\documentclass[11pt,twoside]{article}  
\usepackage{apn3conf}
\def\degr{\hbox{$^\circ$}}
\def\arcmin{\hbox{$^\prime$}}
\def\farcs{\hbox{$.\!\!^{\prime\prime}$}}

\begin{document}
\title{The Post-AGB Star IRAS 16594$-$4656}
\author{Griet C. Van de Steene}
\affil{Royal Observatory of Belgium, Ringlaan 3, 1180 Brussels, Belgium}
\author{Peter A. M. van Hoof}
\affil{APS Division, Physics Dept., Queen's University Belfast, BT7~1NN, Northern Ireland}
\contact{Griet C. Van de Steene}
\email{gsteene@oma.be}
\paindex{Van de Steene, G. C.}
\aindex{van Hoof, P. A. M.}
\authormark{Van de Steene \& van Hoof}

\begin{abstract}
We present the basic properties of the multipolar post-AGB star IRAS
16594$-$4656, and discuss in particular its near infrared spectrum which shows
shock excited H$_2$ and [Fe\,{\sc ii}] emission lines.
\end{abstract}

\section{General Properties}

\begin{figure}
\epsscale{.4}
\plotone{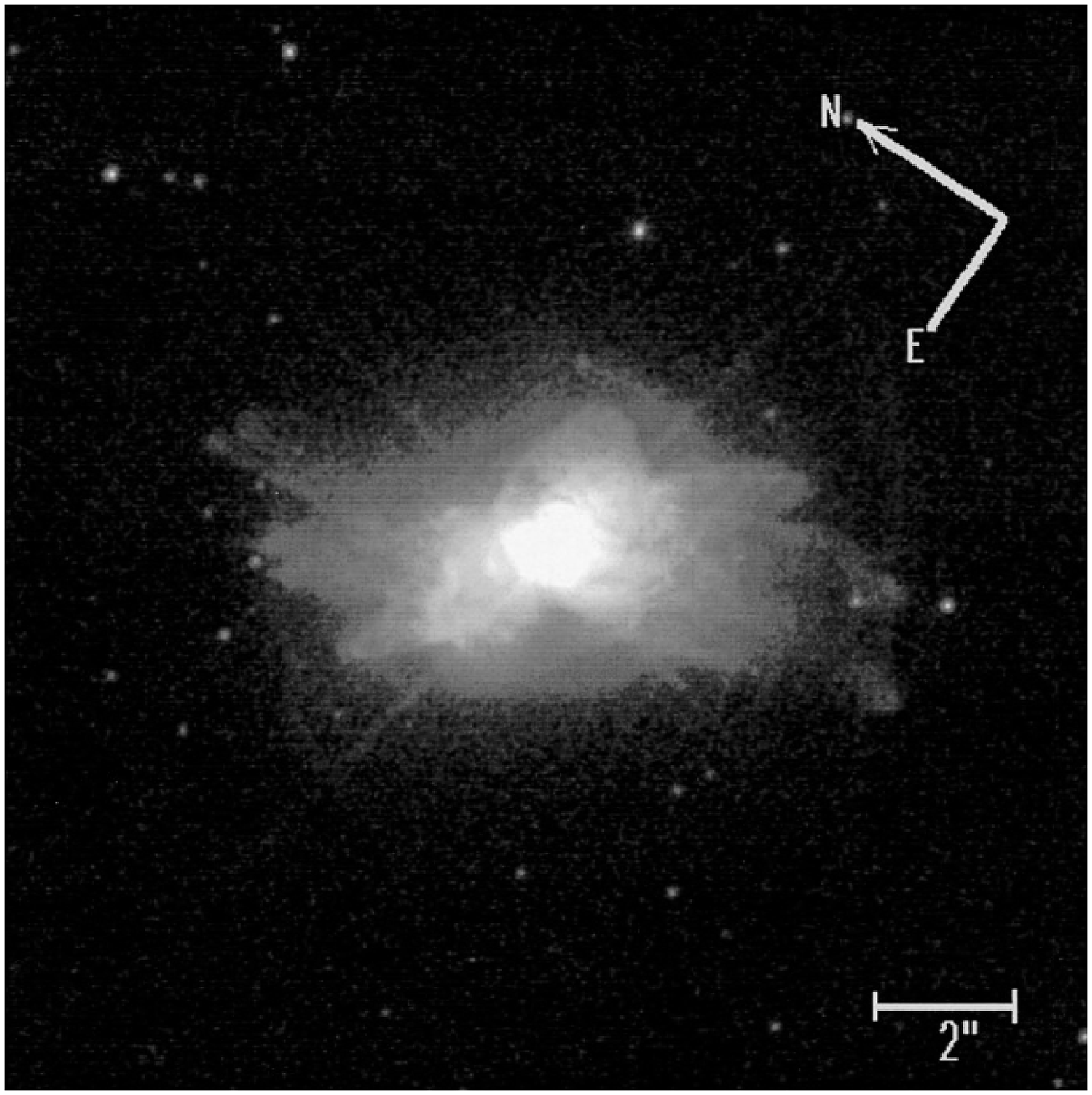}
\hfill
\epsscale{.55}
\plotone{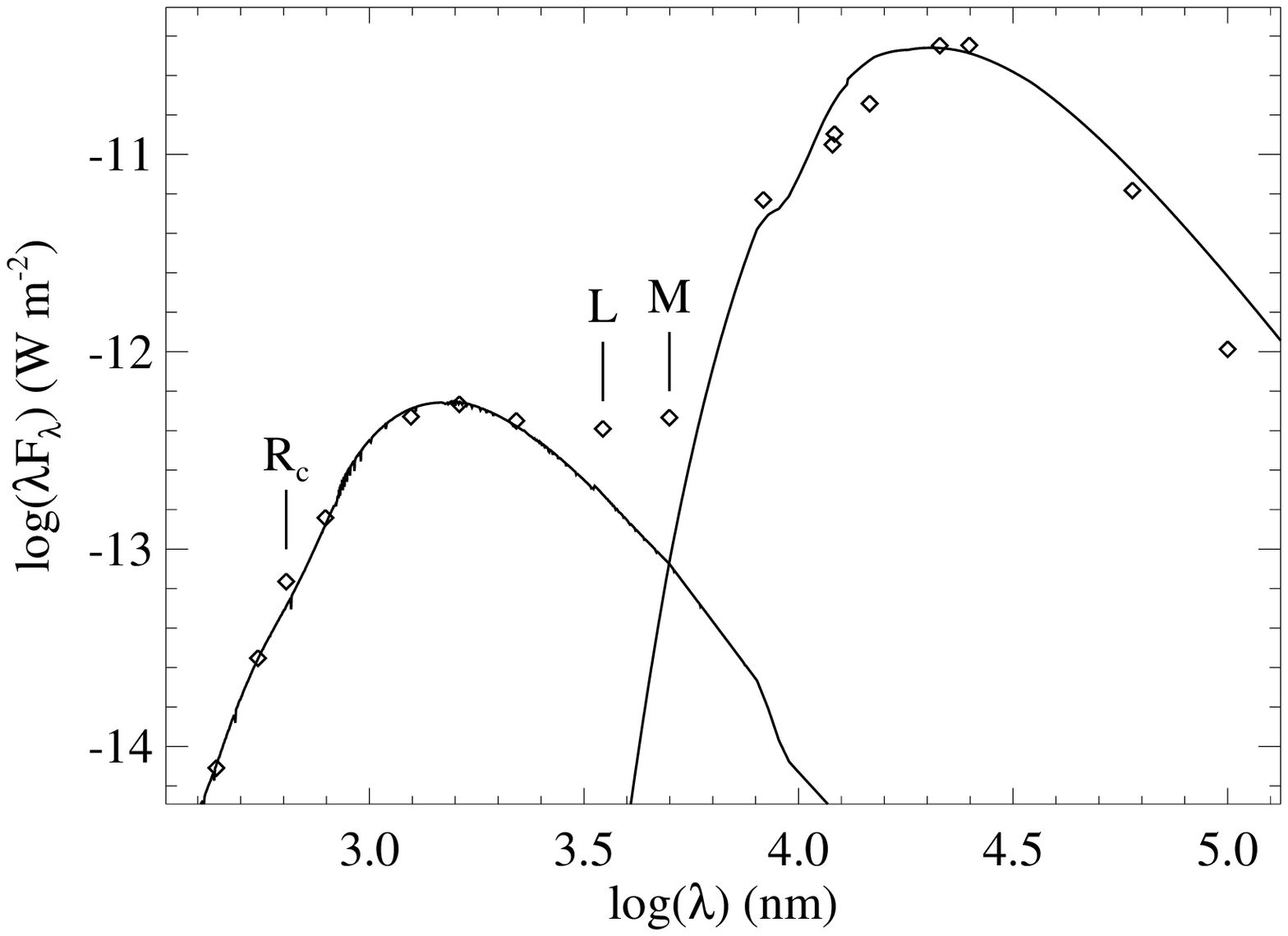}
\vspace{0.3cm}
\caption{The HST image of IRAS 16594$-$4656 (Hrivnak et al. 1999) and its 
spectral energy distribution.}
\label{imaSED}
\end{figure}

IRAS 16594$-$4656 is an optically visible post-AGB star, listed in the {\it
USNO} B1.0 catalog with a position $\alpha$ = 17$^h$ 03$^m$ 10$.\!\!^s$05,
$\delta$ = $-$47\degr\ 00\arcmin\ 27\farcs6 (J2000). In Fig.~\ref{imaSED} we
show the spectral energy distribution (SED) of this star. It has a double
peaked spectrum typical for post-AGB stars. Overplotted is a Kurucz model in
the optical and a black body fit of 188~K in the infrared. The spectrum that
is shown is significantly reddened by interstellar extinction with $A_V = 7.5
\pm 0.4$~mag and $R_V = 4.2$ (Van de Steene \& van Hoof 2003). The excess
emission in the {\it R$_{\rm c}$}-band is due to the strong H$\alpha$
emission. The terminal stellar wind velocity as determined from this P-Cygni
type H$\alpha$ emission is about 126~km\,s$^{-1}$ (Van de Steene et al. 2000b).
The SED clearly shows excess emission in the {\it L}- and {\it M}-bands as
well, possibly indicating the presence of hot dust. The central star has
spectral type B7 (Van de Steene et al. 2000b), indicating a temperature
$T_{\rm eff}$ $=$ 13,000 $\pm$ 1000~K and $\log (g/{\rm cm\,s}^{-2}) = 2$ (Van
de Steene \& van Hoof 2003; Reyniers 2003). The central star is not hot enough
to significantly ionize the surrounding AGB shell. This is corroborated by the
absence of typical planetary nebula lines in the optical and near-infrared
spectrum of this object, and the fact that it has not been detected in the
radio (Van de Steene et al. 2000a). The distance is about $2.2 \pm 0.4$~kpc,
assuming a luminosity of 10,000~L$_\odot$ (Van de Steene \& van Hoof 2003).
This is in good agreement with the distance determination by Su et al. (2001)
of 1.9~kpc assuming a luminosity of 6000~L$_\odot$.

Fig. \ref{imaSED} shows an HST picture of IRAS 16594$-$4656 through the F606W
filter (Hrivnak et al. 1999). It has a multipolar reflection nebula inclined
at intermediate orientation (Su et al. 2001), an elliptical halo of
12.3\arcsec x 8.8\arcsec, and arcs. Its CO expansion velocity is at least
16~km\,s$^{-1}$ (Loup et al. 1990). The ISO spectrum showed that the nebula
has a C-rich chemistry (Garc\'\i a-Lario et al. 1999), although tentative
indications for the presence of crystalline silicates were found.

\section{H$_2$ Emission}

We obtained a full {\it JHK}-band near infrared spectrum of IRAS 16594$-$4656 with {\it SOFI} at
the NTT (ESO). We detected numerous H$_2$ lines. The strongest H$_2$ emission
is seen from (1,0)\,Q(1), (1,0)\,S(1), and (1,0)\,Q(3). The weakest line is
(2,1)\,S(1), and no other lines with $v=2$ or higher have been detected.

\begin{figure}
\begin{minipage}{7cm}
\plotone{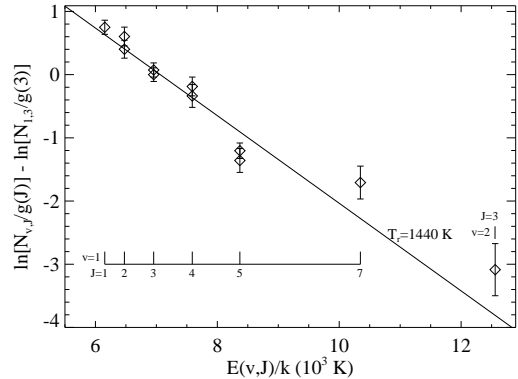}
\end{minipage}
\hfill
\begin{minipage}{7cm}
\caption{The column densities for the various observed ro-vibrational
levels of H$_2$. The
$y$-axis shows the log of the relative column density of molecular
hydrogen in a given state, $N(v,J)$, divided by the statistical
weight, $g(J)$.  The best-fit line based on a rotational
temperature of $T_{\rm r} = 1440$~K is shown as well.}
\end{minipage}
\label{H2plot}
\end{figure}

There are two likely excitation mechanisms for H$_2$ in post-AGB stars: UV
pumping by stellar photons and thermal excitation. We analyzed the H$_2$ lines
to derive the excitation mechanism. More details can be found in Van de Steene
\& van Hoof (2003). Fig.~\ref{H2plot} provides a graphical summary of this
analysis. Comparison of the population of upper levels with different
rotational quantum numbers ($J$-values), but identical vibrational quantum
numbers ($v$-values), provides an estimate for the rotational temperature:
$T_{\rm r} = 1440 \pm 80$~K. The vibrational temperature is measured from the
slope of a line passing through data points with different vibrational quantum
numbers but the same rotational quantum number. This yields $T_{\rm vib} =
1820 \pm 240$~K. The fact that the two values differ only by 1.6~$\sigma$ is
consistent with the assumption that in H$_2$ is mainly collisionally excited.
We also determined the ortho-to-para ratio of molecular hydrogen. This is the
ratio of the total column density of ortho-H$_{2}$ (all odd $J$ states) to
para-H$_2$ (all even $J$ states). We found a ratio of 2.77 $\pm$ 0.19, in good
agreement with the expected ratio of 3 for collisionally excited molecular
hydrogen. A simple model showed that it is unlikely that H$_2$ is thermally
excited by UV heated gas. Hence H$_2$ must be shock excited in IRAS
16594$-$4656.

The ratio H$_2$ (1,0)\,S(1) to Br$\gamma$ is 8.4 after correction for
extinction. Molecular lines produced by collisional excitation of H$_2$ will
be strongest if the collisions are not energetic enough to dissociate H$_2$
and lower its abundance (Hollenbach \& McKee 1989). Such conditions exist in
C-shocks because the gas is heated gradually and remains molecular. Therefore
the strong H$_2$ emission in IRAS~16594$-$4656 argues in favor of a C-shock.
C-shock models of Le Bourlot et al. (2002) indicate that H$_2$ is excited by
shocks with a velocity of 30~km\,s$^{-1}$ in material with a density of
10$^3$~cm$^{-3}$.

\begin{figure}
\epsscale{.45}
\plotone{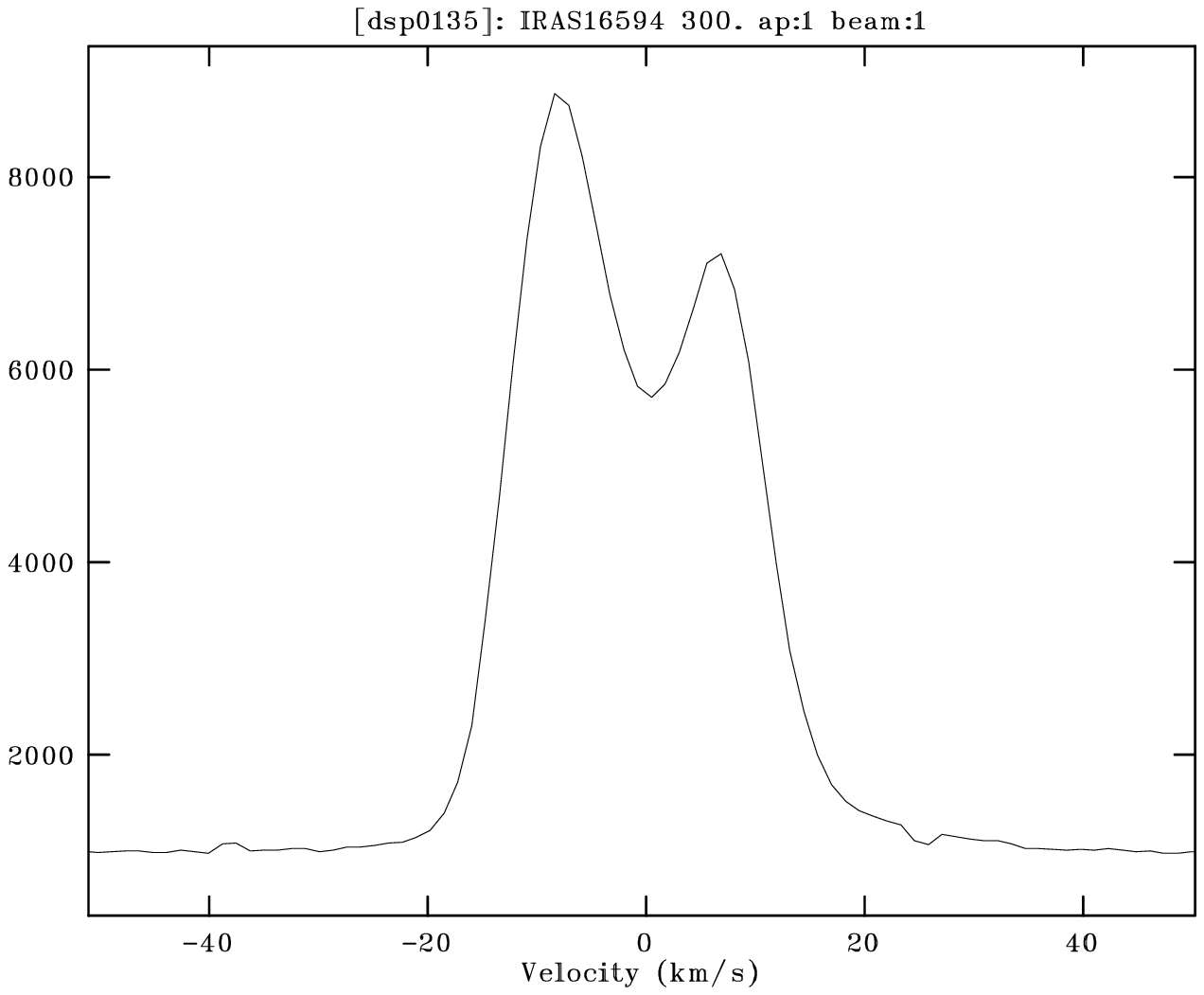}
\hfill
\epsscale{.45}
\plotone{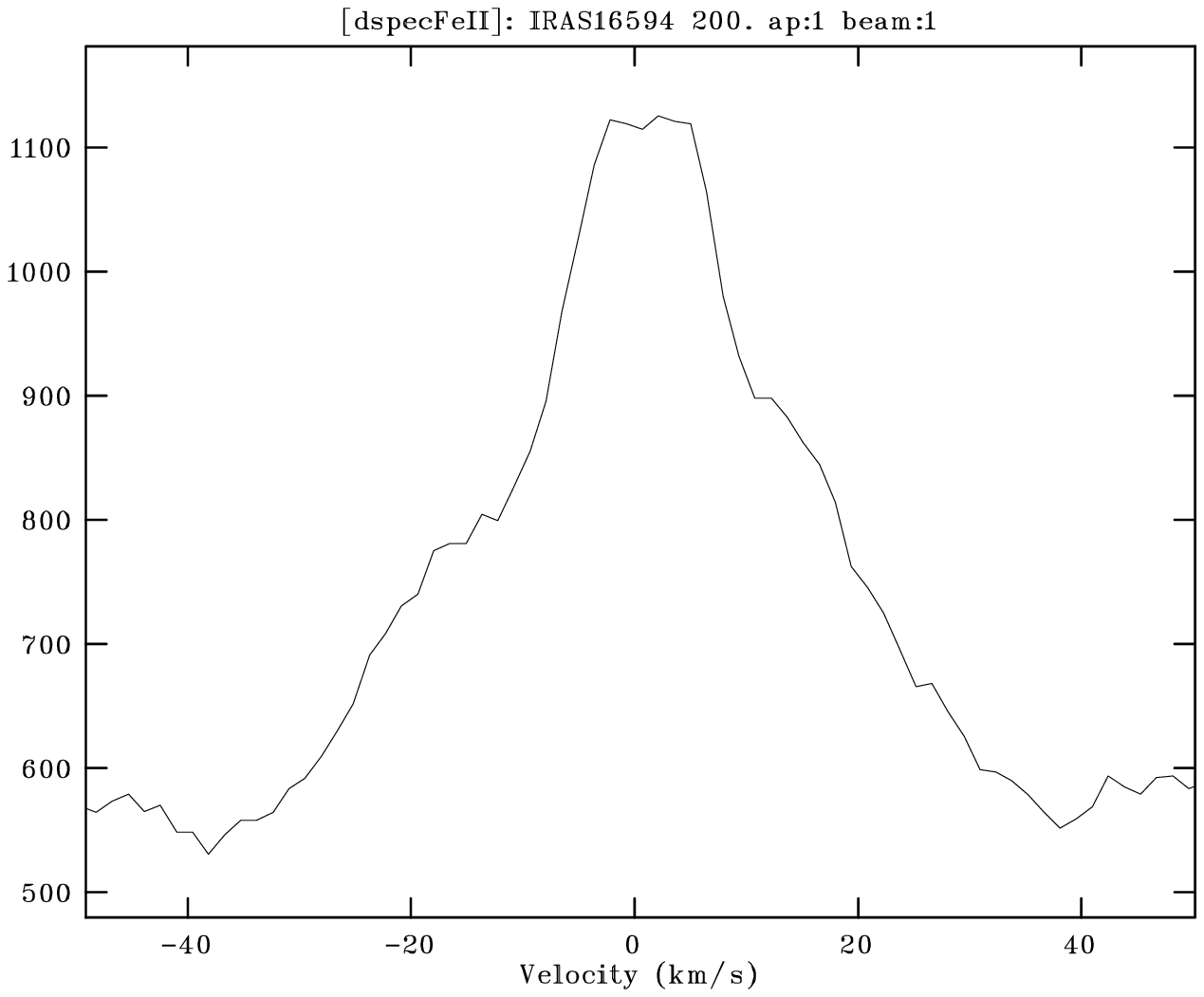}
\caption{The H$_2$ and [Fe\,{\sc ii}] {\it PHOENIX} spectra at 7 deg PA.}
\label{H2FeII}
\end{figure}

In order to investigate the velocity structure and extent of the H$_2$
emission in IRAS 16594$-$4656 we obtained spectra with {\it PHOENIX} on Gemini
South at 3 position angles. The H$_2$ emission is spatially extended. The velocity
difference between the peaks is 15~km\,s$^{-1}$ (Fig. \ref{H2FeII}).

\section{[Fe\,{\sc ii}] 1.644~$\mu$m Emission}

The [Fe\,{\sc ii}] 1.644~$\mu$m line was also detected in the {\it SOFI} spectrum.
The [Fe\,{\sc ii}] $a\,^4\!F$--$a\,^4\!D$ 1.644~$\mu$m over Br$\gamma$
intensity ratio is often used as an indicator of shock excitation. The ratio
expected for shock-excited gas is much larger than 1, but the [Fe\,{\sc ii}]
1.644~$\mu$m/Br$\gamma$ ratio expected for radiatively excited gas is only
approximately 0.06 (Graham et al. 1987). The ratio which we obtained for the
dereddened lines is 0.9, which is much larger than the value typically found
in H\,{\sc ii} regions. Hence the [Fe\,{\sc ii}] emission is possibly shock excited.
However, because the ionization potential of neutral iron is 7.87~eV, and the
dissociation energy of H$_2$ is 4.48~eV (Graham et al. 1987), in principle
H$_2$ and Fe$^+$ cannot coexist in the same region in substantial quantities.
Therefore the H$_2$ and [Fe\,{\sc ii}] emission must originate from different
regions.

The {\it PHOENIX} spectrum shows that the [Fe\,{\sc ii}] emission is not extended.
Hence [Fe\,{\sc ii}] must originate close to the central star, possibly in the
post-AGB wind itself. Shock waves induced by stellar pulsations were proposed
to explain the [Fe\,{\sc ii}] emission in Mira variables (Richter et al.
2003). Post-AGB stars are of course hotter than Miras, but they are usually
variable and still pulsating. Fokin et al. (2001) argue that stellar pulsations
are forming shocks in the atmosphere of the post-AGB star HD 56126. Miras are
believed to have very high mass loss rates, but the mass loss rate in post-AGB
stars may still be quite large as well (up to
10$^{-5}$-10$^{-6}$~M$_\odot$\,yr$^{-1}$, Gauba et al. 2003), as may be
indicated by the strong P-Cygni Balmer lines in the optical spectrum of
IRAS~16594$-$4656 (Van de Steene et al. 2000b). Moreover, post-AGB stars have
higher wind velocities, which lead to the same normalized [Fe\,{\sc ii}] peak
fluxes at lower pre-shock densities. Alternatively, shocked [Fe\,{\sc ii}]
could occur where matter transferred from a binary companion hits an accretion
disk. At this stage there is no corroborating evidence that either a binary
companion or an accretion disk exists in IRAS 16594$-$4656, although the
excess emission in the {\it L}- and {\it M}-bands would be consistent with
such an interpretation.

\section{Conclusions}

IRAS 16594$-$4656 is a multipolar nebula of which the B-type central
star is optically visible. Analysis of the near-infrared spectrum of
this object shows the presence of shock excited emission, but no
photo-ionization. This object gives us the unique opportunity to
study wind-nebula interaction, yet uncompromised by ionization of the
circumstellar shell.

\end{document}